\documentclass[ 
    ,final            % use final for the camera ready runs  
    ,numberedheadings % uncomment this option for numbered sections 
    ,cmfonts          % add further options here if necessary 
  ] 
  {aipproc} 
\usepackage{graphicx,epsfig,slashed} 
 
\layoutstyle{6x9} 
 
\newcommand{\dslash}{{\partial\!\!\!\!/}}     
\newcommand{\eg}{{, {\it e.g.}, }} 

\newcommand{\etal}{{\it et al. }}

\begin{document} 
 
\title{Single-flavor CSL phase in compact stars} 
 
\classification{12.38.Mh, 11.10.St, 12.38.Lg} 
\keywords{Quark Gluon Plasma, Nonperturbative Models, Color Superconductivity} 

\author{David~Blaschke}{ 
  address={Instytut Fizyki Teoretycznej,  
Uniwersytet Wroc{\l}awski, 50-204 Wroc{\l}aw, Poland}, 
altaddress={Bogoliubov Laboratory for Theoretical Physics, 
JINR, 141980 Dubna, Russia} 
} 

\author{Fredrik~Sandin}{ 
  address={IFPA, D\'epartement AGO, Universit\'e de Li\`ege, 4000 Li\`ege, Belgium}, 
altaddress={Division of Physics,  
Lule{\aa} University of Technology, SE-97187 Lule\aa , Sweden} 
} 

\author{Thomas~Kl\"ahn}{ 
  address={Theory Division, Argonne National Laboratory, Argonne IL, USA  
} 
} 

\author{Jens~Berdermann}{ 
  address={DESY Zeuthen, Platanenallee 6, D-15738 Berlin, Germany 
\\ 
E-mail:  
Blaschke@ift.uni.wroc.pl,  
Fredrik.Sandin@gmail.com,  
Thomas.Klaehn@googlemail.com,
Jens.Berdermann@googlemail.com}
} 

\begin{abstract} 
We suggest a scenario where the three light quark flavors are sequentially 
deconfined under increasing pressure in cold asymmetric nuclear matter as\eg
in neutron stars.  
The basis for our analysis is a chiral quark matter model of 
Nambu--Jona-Lasinio (NJL) type with diquark pairing in the spin-1 single 
flavor (CSL), spin-0 two flavor (2SC) and three flavor (CFL) channels. 
We find that nucleon dissociation sets in at about the saturation density, 
$n_0$, when the down-quark Fermi sea is populated (d-quark dripline) due to 
the flavor asymmetry induced by $\beta$-equilibrium and charge neutrality. 
At about $3n_0$ u-quarks appear and a two-flavor color superconducting 
(2SC) phase is formed.
The s-quark Fermi sea is populated only at still higher baryon density, when
the quark chemical potential is of the order of the dynamically generated
strange quark mass.
We construct two different hybrid equations of state (EoS) using the
Dirac-Brueckner Hartree-Fock (DBHF) approach and the EoS by Shen \etal in
the nuclear matter sector.
The corresponding hybrid star sequences have maximum masses of, respectively,
2.1 and 2.0 M$_\odot$. 
Two- and three-flavor quark-matter phases exist only in gravitationally 
unstable hybrid star solutions in the DBHF case, while the Shen-based EoS
produce stable configurations with a 2SC phase component in the core of
massive stars.
Nucleon dissociation due to d-quark drip at the crust-core boundary fulfills 
basic criteria for a deep crustal heating process which is required to 
explain superbusts as well as cooling of X-ray transients.
\end{abstract} 
 
\maketitle 
 
%%%%%%%%%%%%%%%%%%%%%%%%%%%%%%%%%%%%%%%%%%%% 
%% MAINMATTER 
%%%%%%%%%%%%%%%%%%%%%%%%%%%%%%%%%%%%%%%%%%%% 
 
\section{Introduction} 
 
The phenomenology of compact stars is intimately connected to the
EoS of matter at densities well beyond the nuclear
saturation density, $n_0=0.16$ fm$^{-3}$.
Compact stars are therefore natural laboratories for the exploration of 
baryonic matter under extreme conditions, complementary to those created
in terrestrial experiments with atomic nuclei and heavy-ion collisions.
Recent results derived from observations of compact stars provide
serious constraints on the nuclear EoS, see \cite{Klahn:2006ir} and
references therein.
A stiff EoS at high density is needed to explain the high compact-star
masses, $M\sim 2.0$~M$_\odot$, reported for some low-mass X-ray binaries
(LMXBs)\eg 4U 1636-536 \cite{Barret:2005wd}, and the large radius,
$R > 12$~km, of the isolated neutron star RX J1856.5-3754 (shorthand:
RX J1856) \cite{Trumper:2003we}.
Another example is EXO 0748-676, an LMXB for which the compact-star mass
{\it and} radius have been constrained to $M\ge 2.10\pm 0.28$~M$_\odot$
and $R \ge 13.8 \pm 0.18$~km \cite{Ozel:2006km}.   
However, the status of the results for the latter object is unclear,
because the gravitational redshift $z=0.35$ observed in the X-ray burst
spectra \cite{Cottam:2002} has not been confirmed, despite numerous attempts.
While compact-star phenomenology apparently points towards a stiff EoS
at high density, heavy-ion collision data for kaon production 
\cite{Fuchs:2005zg}
and elliptic flow \cite{Danielewicz:2002pu} set an upper limit on the
stiffness of the EoS \cite{Klahn:2006ir}.

A key question regarding the structure of matter at high density is  
whether a phase transition to quark matter occurs inside compact stars,
and whether it is accompanied by unambiguous observable signatures.
It has been argued  that the observation of a compact star with high mass and 
large radius, likewise reported for EXO 0748-676,
would be incompatible with a quark core \cite{Ozel:2006km}, because the 
softening of  the EoS due to quark deconfinement lowers the maximum mass and
the radius in comparison to the nuclear matter case. 
However, Alford \etal \cite{Alford:2006vz} have demonstrated with a few 
counter examples that quark matter cannot be excluded by this argument. 
In particular, for a recently developed hybrid star EoS \cite{Klahn:2006iw},
based on the DBHF approach in the nuclear sector and a three-flavor chiral 
quark model \cite{Blaschke:2005uj}, stable hybrid stars in the mass range 
from 1.2 M$_\odot$ up to 2.1 M$_\odot$ have been obtained, in accordance with 
modern mass-radius constraints, see also \cite{Blaschke:2007ri}. 
In that model, a sufficiently low critical density for quark deconfinement
was achieved with a strong diquark coupling, while a sufficient stiffness for 
a high maximum mass of the compact star sequence was obtained with a repulsive 
vector meanfield in the quark matter sector.
The corresponding hybrid EoS for symmetric matter was shown to fulfill 
the constraints from elliptic flow data in heavy-ion collisions.
%The maximum compact star mass was delimited  
%by the onset of the strange quark matter in the  
%CFL phase which entailed a softening of the EoS and rendered the hybrid star 
%unstable against gravitational collapse.  
In the present work we want to discuss a new scenario of quark deconfinement, 
which could play an important role in asymmetric matter, in particular for
the phenomenology of compact stars.

\begin{figure} %[ht] 
\begin{tabular}{ll} 
\includegraphics[angle=0,width=0.5\textwidth]{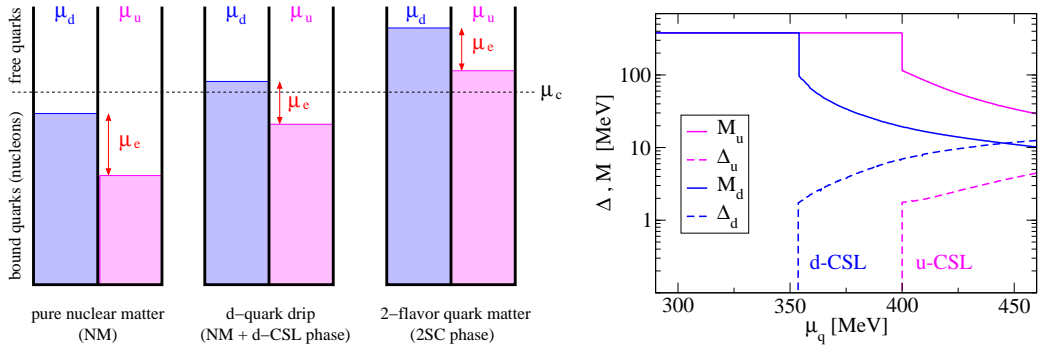}& 
\includegraphics[angle=0,width=0.4\textwidth,clip=]{CSL-gaps.eps} 
\end{tabular} 
\caption{Left panel: Chemical potentials of up and down quarks (strange quark
sector not shown). With increasing quark chemical potential 
$\mu_q=(\mu_u+\mu_d)/2$ in isospin asymmetric matter the quark flavors pass 
sequentially the threshold ($\mu_c$) for chiral symmetry restoration 
(deconfinement), which entails nucleon dissociation. 
Right panel: Solution of the NJL gap equations under isospin asymmetry.} 
    \label{f:phases} 
\end{figure}

Chiral quark models of the NJL type with dynamical chiral symmetry breaking 
have the property that the restoration of this symmetry (and the related quark 
deconfinement) at zero temperature is flavor specific. 
When solving the gap and charge-neutrality equations
selfconsistently one finds that the chiral symmetry restoration 
for a given flavor occurs when the chemical potential of that flavor reaches
a critical value that is approximately equal to the dynamically generated 
quark mass, $\mu_{f}=\mu_c \approx m_{f}$, where $f=u, d, s$. 
In asymmetric matter the quark chemical potentials are different. 
Consequently, the NJL model behavior suggests that the critical density of
deconfinement is flavor dependent, see Fig. \ref{f:phases}.
In this approach the down quark flavor is the first to
drip out of nucleons as the density increases, followed by the
up quark flavor and eventually also by strange quarks.
This behavior is absent in simple and commonly applied thermodynamic 
bag model equations of state since they are essentially flavor blind.

Under the $\beta$-equilibrium condition in compact stars 
the chemical potentials of quarks and electrons are related by 
$\mu_d=\mu_s$ and $\mu_d=\mu_u+\mu_e$. 
The mass difference between the strange and the light quark flavors  
$m_s \gg m_u, m_d$ has two consequences: (1) the down and strange 
quark densities are different, so charge neutrality requires a 
finite electron density and, consequently, (2) $\mu_d>\mu_u$. 
When increasing the baryochemical potential, the d-quark chemical
potential is therefore the first to reach the critical value $\mu_c$ 
where the chiral symmetry gets (approximately) restored in a first-order
transition and a finite density of d-quarks appears. 
Due to the finite value of $\mu_e$, the u-quark chemical potential  
is still below $\mu_c$ while the s-quark density is zero due to 
the high s-quark mass. A {\it single-flavor} d-quark phase
therefore forms in co-existence with the positively charged
nuclear-matter medium.

Why has this interesting scenario been left unnoticed? 
On the one hand, bag models, which are commonly used to describe quark 
matter in compact star interiors cannot address sequential deconfinement. 
On the other hand, the single-flavor d-quark phase is negatively charged
and cannot be neutralized in a purely leptonic background. 
This was a reason to disregard it in dynamical approaches like the NJL models.
In the following we discuss the single-flavor phase for the first time under
the natural assumption that the neutralizing background is nuclear matter.
Since nucleons are bound states of quarks, the physical context in which such a
mixed phase of nucleons and free d-quarks occurs is that of the dissociation 
of nucleonic bound states of quarks (Mott effect). 

\section{Phase transition to quark matter: nucleon dissociation} 

The task to develop a unified description of the phase transition from 
nuclear matter to quark matter on the quark level, as a dissociation of 
three-quark bound states into their constituents in the spirit of a Mott 
transition has not been solved yet. 
Only some aspects of such a description have been revealed within a 
nonrelativistic potential model \cite{Horowitz:1985tx,Ropke:1986qs} and 
within the NJL model \cite{Lawley:2006ps}.
We will consider a chemical equilibrium of the type
$n+n \leftrightarrow p + 3 d$, which results in a mixed phase of nucleons
and down quarks once the d-quark chemical potential exceeds the critical value.
This scenario is analogous to the dissociation of nuclear clusters in the
crust of neutron stars (neutron dripline) and the effect may therefore
be called the {\em d-quark dripline}.
We approximate the quark and nucleon components as subphases described by 
separate models. 
For the nuclear matter subphase we use two alternatives: 
(1) the DBHF approach 
\cite{deJong:1997hr,boelting99,honnef,DaFuFae04,DaFuFae05} 
 with the relativistic Bonn~A potential, where the nucleon selfenergies are 
based on a T-matrix  obtained from the Bethe-Salpeter equation in the ladder 
approximation and 
(2) the EoS by Shen \etal \cite{Shen:1998gq}, which is based on a relativistic 
mean field theory  and includes the contribution of heavy nuclei, described 
within the Thomas-Fermi approximation. 
The quark matter phase is described within a three-flavor NJL-type  
model including diquark pairing channels 
\cite{Blaschke:2005uj,Ruster:2005jc,Abuki:2005ms,Warringa:2005jh}. 
\begin{figure}
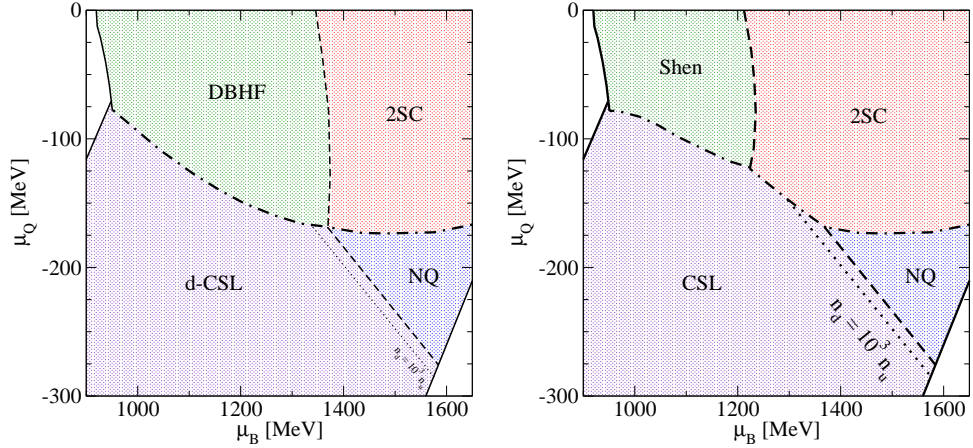
 %[ht] 
\begin{tabular}{ll} 
\includegraphics[angle=0,height=0.4\textwidth]{DBHF-NJL_phasediag.eps}& 
\includegraphics[angle=0,height=0.4\textwidth]{Shen-NJL_phasediag.eps} 
\end{tabular} 
\caption{Phase diagram in the plane of baryon and charge chemical  
potential. The dash-dotted line denote the border between oppositely charged  
phases. The nuclear matter EoS is DBHF (left panel) and Shen \etal 
(right panel).} 
    \label{f:phasediag} 
\end{figure} 
The path-integral representation of the partition function is given by
\begin{eqnarray}   
\label{Z}  
Z(T,\hat{\mu})&=&\int {\mathcal D}\bar{q}{\mathcal D}q   
\exp \left\{\int_0^\beta d\tau\int d^3x\,\left[   
        \bar{q}\left(i\dslash-\hat{m}+\hat{\mu}\gamma^0\right)q+  
{\mathcal L}_{\rm int}   
\right]\right\},  
\end{eqnarray}   
\begin{eqnarray}   
\label{Lint}  
{\mathcal L}_{\rm int} &=& G_S\bigg\{ 
        \sum_{a=0}^8\big[(\bar{q}\tau_aq)^2 + (\bar{q}i\gamma_5\tau_aq)^2\big]
        +\eta_{D0}\sum_{A=2,5,7} j_{D0,A}^\dagger  j_{D0,A} 
        +\eta_{D1}~j_{D1}^\dagger j_{D1} \bigg\},   
\end{eqnarray}   
where $\hat{\mu}=\frac{1}{3}\mu_B+{\rm diag}_f(\frac{2}{3},-\frac{1}{3},-\frac{1}{3})\mu_Q+\lambda_3\mu_3+\lambda_8\mu_8$ 
is the diagonal quark chemical potential matrix  
and $\hat{m}={\rm diag}_f(m_u,m_d,m_s)$ is 
the current-quark mass matrix.  
For $a=0$, $\tau_0=(2/3)^{1/2}{\mathbf 1}_f$, otherwise $\tau_a$ and   
$\lambda_a$  are Gell-Mann matrices acting in flavor and color spaces,  
respectively.  
$C=i\gamma^2\gamma^0$ is the charge conjugation operator and   
$\bar{q}=q^\dagger\gamma^0$.  
The scalar quark-antiquark current-current interaction is given  
explicitely and has coupling strength $G_S$. The 3-momentum cutoff, 
$\Lambda$,  is fixed by low-energy QCD phenomenology (see table I of 
\cite{Grigorian:2006qe}). 
The spin-0 and spin-1 diquark currents are 
$j_{D0,A}=q^TiC\gamma_5\tau_A\lambda_Aq$ and  
$j_{D1}=q^TiC(\gamma_1\lambda_7+\gamma_2\lambda_5+\gamma_3\lambda_2)q$. 
While the relative coupling strengths $\eta_{D0}$ and $\eta_{D1}$  
are essentially free parameters,  we restrict the discussion 
to the Fierz values, $\eta_{D0}=3/4$ and $\eta_{D1}=3/8$, see  
\cite{Buballa:2003qv}.
Color superconducting phases in QCD with one flavor have first been discussed 
in Refs. \cite{Schafer:2000tw,Alford:2002rz,Schmitt:2004et}, where also the
special role of the spin-1 color-spin locking phase has been pointed out that
here all quarks participate in the pairing with a gap of the order of 1 MeV or
even below. This feature of the CSL phase is robust, as was demonstrated for
our above isotropic ansatz for the spin-1 diquark current introduced in 
\cite{Aguilera:2005tg}, and for its generalizations to the nonlocal
case \cite{Aguilera:2006cj} and to a selfconsistent Dyson-Schwinger approach
\cite{Marhauser:2006hy}. 

The gaps and the renormalized masses are determined by minimization of the 
mean-field thermodynamic potential, under the constraints of charge neutrality
and $\beta$-equilibrium. For further details, see
\cite{Blaschke:2005uj,Ruster:2005jc,Abuki:2005ms,Warringa:2005jh}. 
In Fig. \ref{f:phasediag} we plot the thermodynamically favored phase in
the plane of baryon and charge chemical potentials.
 
\begin{figure}
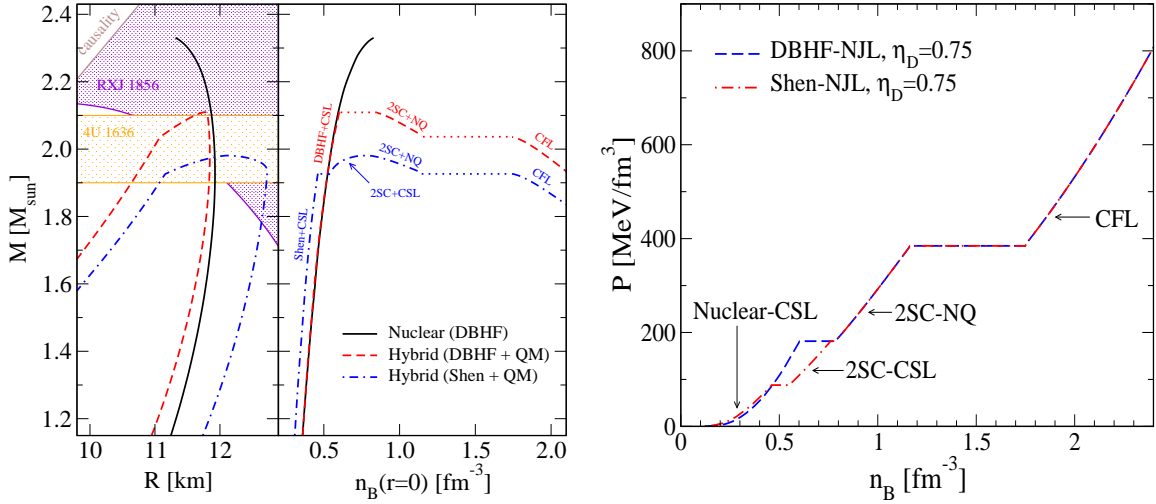
 %[ht] 
\begin{tabular}{ll} 
\includegraphics[height=0.45\textwidth,clip=]{sequences4.eps} &
\includegraphics[height=0.45\textwidth,width=0.5\textwidth,clip=]{P_of_nB.eps} 
\end{tabular} 
\caption{Left panel: 
Compact star sequences. The phase structure of the core changes with  
increasing density, as indicated in the figure. Constraints on the mass come 
from 4U 1636 \cite{Barret:2005wd} and on the mass-radius relation from  
RX J1856 \cite{Trumper:2003we}.
Right panel: Hybrid equations of state used in the calculation of the
compact star sequences.
} 
\label{f:eos-config} 
\end{figure}

The hybrid EoS corresponds to the dash-dotted lines in Fig. \ref{f:phasediag}
and are constructed such that the mixture of nuclear matter, quark matter and
leptons is charge neutral. Using the hybrid EoS we calculate the corresponding
compact star sequences by solving the Oppenheimer-Volkoff equations for
hydrostatic equilibrium. The hybrid-star sequences fulfill all modern
constraints on the mass-radius relationship, see Fig. \ref{f:eos-config}.  
For the DBHF hybrid EoS all stars with DBHF+CSL matter in the
core are stable equilibrium solutions, while the appearance of
u-quarks and the associated formation of a 2SC subphase renders
the sequence unstable. The situation is somewhat different for
the Shen hybrid EoS, because in addition to Shen+CSL stars there
are stable solutions with 2SC+CSL matter in the core. In both
cases configurations with strange quarks in the core are unstable.

The hybrid star sequences 'masquerade' as neutron stars
\cite{Alford:2004pf}, because the mechanical properties are
similar to those of nuclear matter stars and the transition
from nuclear matter to the mixed phase is associated with
a relatively small discontinuity in the density.
Unmasking neutron star interiors might therefore require
observables based on transport properties, which could be
strongly modified in presence of color superconductivity. 
It has been suggested to base such tests of the structure
of matter at high density on an analysis of the cooling
behavior \cite{Blaschke:2006gd,Popov:2004ey,Popov:2005xa,Grigorian:2006pu} 
or the stability of rapidly spinning stars against r-modes  
\cite{Madsen:1999ci,Drago:2007iy}. 
It has turned out that these phenomena are sensitive to
the details of color superconductivity in quark matter.

%%%%%%%%%%%%%%%%%%%%%%%%Jens%%%%%%%%%%%%%%%%%%%%%%%%%%%%%%%%%%%%%%%%%%%%%%% 
\section{Bulk viscosity and Urca emissivity of the single-flavor CSL 
phase} 
 
According to \cite{Andersson:1997xt} rotating compact stars
would be unstable against r-modes in the absence of viscosity
\cite{Andersson:2000mf}.
Constraints on the composition of compact-star interiors can
therefore be obtained from observations of millisecond pulsars   
\cite{Madsen:1999ci,Drago:2007iy}.    
In such investigations the bulk viscosity is a key quantity.
We therefore consider some relevant aspects here, starting
with the two-flavor color superconducting phases following
the approach described in Ref. \cite{Sa'd:2006qv}.
%Note that the 2SC phase considered in \cite{Alford:2006gy}
%is a three-flavor phase, for which the nonleptonic process
%$u+d \leftrightarrow u+s$ is the dominant contribution.
%This process is irrelevant here.

\begin{figure}%[ht] 
\begin{tabular}{cc}
\includegraphics[height=3cm,width=5cm]{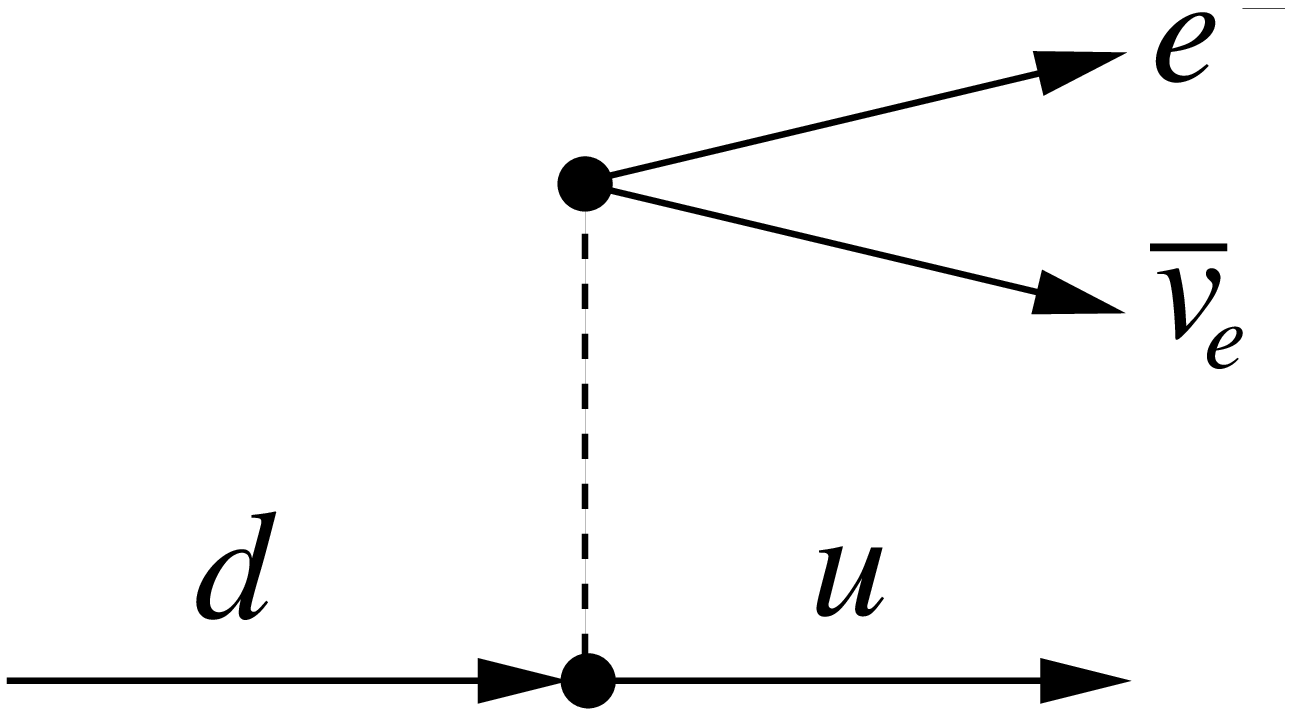}& 
\includegraphics[height=4cm,width=7cm]{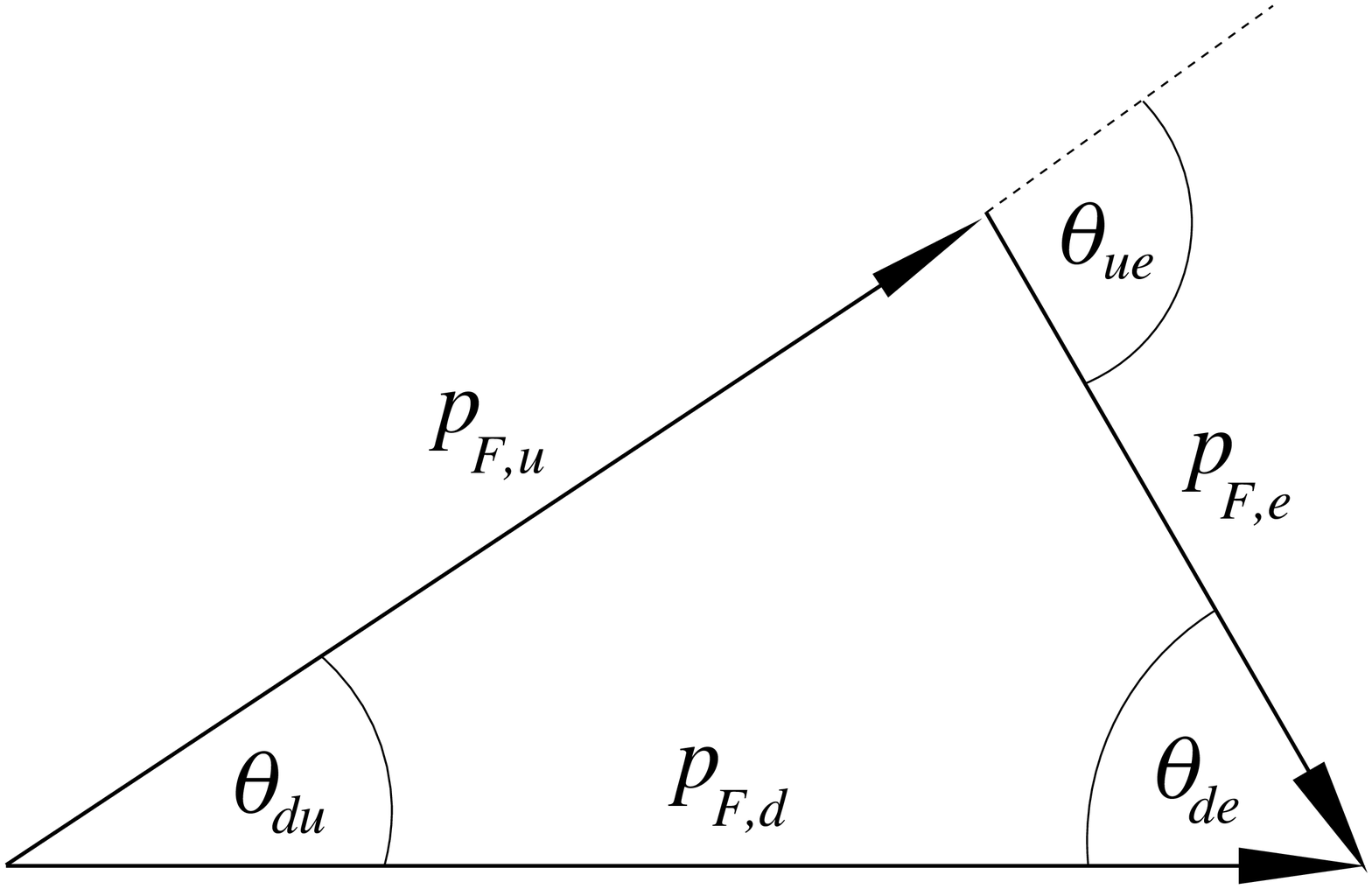} 
\end{tabular}
\caption{\small Direct Urca process in quark matter (left) and triangle
of momentum conservation for it (right).}
\label{f:ptriangle} 
\end{figure}

The temperature-dependent bulk viscosity for the 2SC-CSL phase
has been calculated self-consistently in \cite{Blaschke:2007bv}
and is based on the flavor-changing weak processes of electron
capture and $\beta$ decay 
\begin{eqnarray} 
u + e^- \rightarrow d + \nu_e~~,~~~d \rightarrow u + e^- + \bar{\nu}_e~. 
\end{eqnarray}  
It has been shown that the bulk viscosity is related to the direct
URCA emissivity, which for normal quark matter was first calculated
by Iwamoto \cite{Iwamoto:1982} and can be expressed as 
\begin{equation} 
\label{urca} 
\varepsilon_{0} \simeq  
\frac{914~\pi}{1680}G_F^2~\mu_e \mu_u \mu_d~ T^6~\theta_{ue}^2.  
\end{equation} 
Here $G_F$ is the weak coupling constant  
and $\theta_{ue}$ is the angle between the up-quark and electron  
momenta, which is obtained from momentum conservation in the matrix  
element, see Fig. \ref{f:ptriangle}.
The triangle of momentum conservation holds for the late cooling stage,
when the temperature is below 1 MeV and neutrinos are untrapped. 
Trigonometric relations are used to find an analytical expression
for momentum conservation.
To lowest order in $\theta_{de}$ the result is 
\begin{equation} 
\label{eq0} 
p_{F,d}-p_{F,u}-p_{F,e}\simeq - \frac{1}{2} p_{F,e}~\theta_{de}^2 ~. 
\end{equation} 
For small angles $\theta_{de} \simeq \theta_{ue}$, so it is possible to
obtain an expression for the matrix element of the direct URCA process. 
Following Iwamoto \cite{Iwamoto:1982} one has to account either for 
quark-quark interactions to lowest order in the strong coupling
constant, $\alpha_s$, (\ref{eq1}) or the effect of finite  masses (\ref{eq2}):
\begin{eqnarray} 
\mu_i &=& p_{F,i}\left(1+\frac{2}{3\pi}\alpha_s\right) ~,~~~{i=u,d}  
\label{eq1}\\ 
\mu_i &\simeq&  
p_{F,i}\left[1+\frac{1}{2}\left(\frac{m_i}{p_{F,i}}\right)^2\right]~,  
~~{i=u,d,e}~~.  
\label{eq2} 
\end{eqnarray} 
From (\ref{eq0})-(\ref{eq2}) and the $\beta$-equilibrium condition,  
$\mu_d = \mu_u + \mu_e$, the angle $\theta_{de}$ that determines the
emissivity (\ref{urca}) is obtained
%an expression for the angle  $\theta_{de}$ determining 
%the emissivity  Eq. (\ref{urca}) can be found similar terms for Eq. (\ref{eq0}) 
\begin{eqnarray}\label{betaeq} 
\theta_{de}^2 \simeq \left\{ \begin{array}{cl}  
  \frac{4}{3\pi}\alpha_s \\[0.2cm] 
\frac{m_d^2}{p_{F,e}p_{F,d}} 
\left[1-\left(\frac{m_u}{m_d}\right)^2\left(\frac{p_{F,d}}{p_{F,u}}\right) 
-\left(\frac{m_e}{m_d}\right)^2\left(\frac{p_{F,d}}{p_{F,e}}\right)\right] 
\\ 
\end{array}\right.~. 
\end{eqnarray} 
%The mass effect for current quark masses has been found negligible in 
% comparison to the perturbative quark-quark interaction effect in the region 
% of the QCD phase diagram where the perturbative treatment is possible at all 
% and the quark masses are of the order of the current values $m_{u,d} \sim 5-9~% {\rm MeV}$ \cite{Berdermann:2007mx}.  
If interactions and masses are neglected, or the Fermi sea of one species is
closed as in the single-flavor CSL phase, it follows that the triangle of
momentum concervation in Fig. \ref{f:ptriangle}) degenerates to a line or
can even not be closed. 
In that case the matrix element vanishes with the consequence that the direct 
URCA process does not occur, and also the bulk viscosity is zero. 
However, in the mixed nuclear-CSL phase there could be important friction
and pair-breaking/formation processes, which we have not yet studied in detail.
This could be an interesting issue for further investigation due to the large
difference in the masses of baryons and deconfined quarks.
%we introduced above a mixed nuclear-CSL phase, where due to the 
%d-drip equilibrium, friction and pair-breaking/formation processes can be 
%important. This kind of processes we did not yet estimate, but could be 
%especially interesting due to a large mass-difference between nucleon and 
%quark masses in the momentum conservation. 

\section{Mechanism for deep crustal heating} 

Superbursts are rare, puzzling phenomena observed as a extremely long 
(4-14 hours) and energetic ($\sim ~10^{42} {\rm erg}$) type-I X-ray bursts 
from LMXBs. 
They take place if the accreted hydrogen and helium at the surface burns in 
an unstable manner, which is the normal case \cite{Stejner:2006tj}.   
As suggested by the authors of Ref. \cite{Page:2005ky}, superbursts could 
originate from accreting strange stars with a tiny crust and a core of 
three-flavor quark matter in the superconducting color flavor locked (CFL) 
phase, since it fulfills the constraints on matter properties from their 
superburst scenario.  
Of particular importance for the scenario is that in this phase the neutrino 
emissivity and heat conductivity are suppressed by pairing gaps affecting all 
quark species  \cite{Blaschke:1999qx,Page:2000wt,Blaschke:2000dy}.
The mechanism underlying the superburst phenomenon is unstable thermonuclear 
burning of carbon in the crust at column depths of about 
$10^9~{\rm g~cm^{-2}}$ \cite{Cumming:2001wg}.  
The carbon itself is a remnant of the burning of accreted hydrogen and helium 
at the surface. 
The ignition of observed superburst light curves takes place at a depth 
 where the crust reaches temperatures around $6\times 10^8$ K and column depths
of about $10^{12}~{\rm g~cm^{-2}}$.  
Such high temperatures in the crust at a certain depth are caused by 
deep crustal heating \cite{Haensel:1990,Ushomirsky:2001pd,Shternin:2007md}.    
The important ingredients for the strange star model of \cite{Page:2005ky} are 
a thin baryonic crust of 100 to 400~m thickness, a sufficient energy release 
of 1 to 100 MeV per accreted nucleon by conversion into strange matter, 
a suppression of the fast direct URCA neutrino emissivity to the order of 
 $10^{21}~{\rm~ erg~ cm^{-3}~ s^{-1}}$, and a thermal conductivity, $\kappa$,
of quark matter in the range $10^{19}-10^{22}~{\rm erg~cm^{-1}~s^{-1}~K^{-1}}$.
   
\begin{figure} [!th]
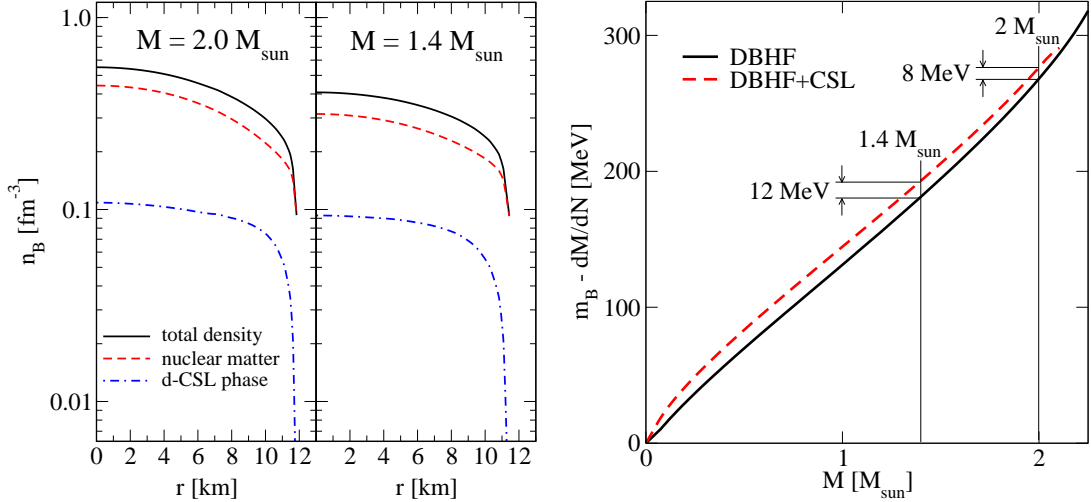
 
\begin{tabular}{ll} 
\includegraphics[angle=0,height=0.45\textwidth]{profiles_DBHF-dCSL.eps}& 
\includegraphics[angle=0,height=0.45\textwidth]{DBHF-CSL_dmdn.eps} 
\end{tabular} 
\caption{Left panel: 
Density profiles of two stars with masses $1.4$~M$_\odot$ and  
$2.0$~M$_\odot$. Note that the mixed phase of d-CSL quark matter with nuclear 
matter extends up to the crust-core boundary. 
Right panel:  
Energy release per nucleon as a function of the compact star mass. 
An upper estimate for the energy release from the conversion of DBHF nuclear 
matter to DBHF-CSL hybrid matter gives 12 (8) MeV for a compact star core 
with mass 1.4 (2.0) $M_\odot$. 
}  
\label{f:profile-release} 
\end{figure}      

One of the main arguments for strange matter is the fact that superconducting 
phases, like the CFL phase, can suppress fast neutrino emission processes of 
all quark flavors and are able to fulfill the fusion ignition condition.
However, as we have shown above, in the single flavor CSL-phase the fast 
direct URCA process is not possible at all, whereas slow neutrino cooling 
processes  like bremsstrahlung of electrons and d-quarks occur.  

As one can see from Fig. \ref{f:profile-release} the energy release from the 
conversion of DBHF nuclear matter to DBHF-CSL hybrid matter gives 12 (8) MeV 
for a compact star core with mass 1.4 (2.0) $M_\odot$, which is in the range
1 -- 100~MeV and could, in principle, explain burst ignition at appropriate
depths for a suitable value of $\kappa$.    
Therefore, the compact star does not necessarily need to be made of strange 
matter, but could be a hybrid star with quark matter in the d-CSL phase and a 
thin crust.     
Stejner {\it et al.} \cite{Stejner:2006tj} show that deep crustal heating 
mechanisms at the crust-core boundary like the conversion of baryonic matter 
to strange quark matter, which can fulfill the constraints of the superburst 
scenario do provide a consistent explanation of the cooling of soft X-ray 
transients too. 
Along the lines of this argument we claim that the d-quark drip effect at the 
crust core boundary, which leads to a mixture of nuclear matter with 
single-flavor quark matter in the CSL phase can serve as a deep crustal 
heating mechanism \cite{Blaschke:2008vh,Blaschke:2008br}. 
Superbursts and the cooling of X-ray transients 
are not only consistent with quark matter in compact stars but may qualify as 
a signature!

\section{Conclusions} 
In this contribution we have suggested a new quark-nuclear hybrid EoS for
compact star applications that fulfills modern observational constraints
from compact stars. 
Due to isospin asymmetry, down-quarks may ``drip out'' from nucleons and  
form a single-flavor color superconducting (CSL) phase that is mixed  
with nuclear matter already at the crust-core boundary in compact stars. 
The CSL phase has interesting cooling and transport properties that are in 
accordance with constraints from the thermal and rotational evolution of  
compact stars.  
It remains to be investigated whether this new compact star composition
could lead to unambiguous observational consequences. 
We conjecture that the d-quark drip may serve as an effective deep crustal 
heating mechanism for the explanation of the puzzling superburst phenomenon 
and the cooling of X-ray transients. 
 
\section*{Acknowledgements} 
We thank the organizers for bringing together colleagues from different 
communities and for creating the stimulating atmosphere at the Workshop in 
Spa, Belgium.
D.B. is supported in part by the Polish Ministry of Science and Higher  
Education, grant No. N~N~202 0953 33; 
T.K. is grateful for partial support from the Department of Energy, 
Office of Nuclear Physics, contract no.\ DE-AC02-06CH11357. 
The work of F.S. was supported by the Belgian fund for scientific research 
(FNRS).
D.B. and F.S. acknowledge the European Science Foundation for supporting their 
participation at the Workshop of the Research Networking Programme ``The 
Physics of Compact Stars (CompStar)'' in L\c{a}dek Zdr\'oj, Poland, where part 
of this work has been done.

\end{document}